\documentclass[twocolumn,table,xcdraw]{aastex63}

\definecolor{auburn}{rgb}{0.43, 0.21, 0.1}
\definecolor{bostonuniversityred}{rgb}{0.8, 0.0, 0.0}
\definecolor{royalazure}{rgb}{0.0, 0.22, 0.66}
\definecolor{myBlue}{HTML}{5480f1}
 \hypersetup{linkcolor=bostonuniversityred,citecolor=royalazure,filecolor=cyan,urlcolor=myBlue}
\usepackage[normalem]{ulem}

\received{date}
\revised{date}
\accepted{date}
\submitjournal{ApJL}

\shorttitle{Hiding signatures of gravitational instability}
\shortauthors{Rowther \& Meru}

\graphicspath{{./}{Images/}}
\usepackage{lipsum}
\begin{document}

\title{Hiding Signatures of Gravitational Instability in Protoplanetary Discs with Planets}

\correspondingauthor{Sahl Rowther}
\email{sahl.rowther@warwick.ac.uk}


\author[0000-0003-4249-4478]{Sahl Rowther}
\affiliation{Centre for Exoplanets and Habitability, University of Warwick, Coventry CV4 7AL, UK}
\affiliation{Department of Physics, University of Warwick, Coventry CV4 7AL, UK}

\author[0000-0002-3984-9496]{Farzana Meru}
\affiliation{Centre for Exoplanets and Habitability, University of Warwick, Coventry CV4 7AL, UK}
\affiliation{Department of Physics, University of Warwick, Coventry CV4 7AL, UK}

\author[0000-0001-6831-7547]{Grant M. Kennedy}
\affiliation{Centre for Exoplanets and Habitability, University of Warwick, Coventry CV4 7AL, UK}
\affiliation{Department of Physics, University of Warwick, Coventry CV4 7AL, UK}
\author[0000-0003-0856-679X]{Rebecca Nealon}
\affiliation{Department of Physics and Astronomy, University of Leicester, Leicester, LE1 7RH, United Kingdom}
%
\author[0000-0001-5907-5179]{Christophe Pinte}
\affiliation{School of Physics and Astronomy, Monash University, Clayton Vic 3800, Australia}



\begin{abstract}
We carry out three dimensional SPH simulations to show that a migrating giant planet strongly suppresses the spiral structure in self-gravitating discs. 
We present mock ALMA continuum observations which show that in the absence of a planet, spiral arms due to gravitational instability are easily observed. Whereas in the presence of a giant planet, the spiral structures are suppressed by the migrating planet resulting in a largely axisymmetric disc with a ring and gap structure. Our modelling of the gas kinematics shows that the planet's presence could be inferred, for example, using optically thin $^{13}$C$^{16}$O.  Our results show that it is not necessary to limit the gas mass of discs by assuming  high dust-to-gas mass ratios in order to explain a lack of spiral features that would otherwise be expected in high mass discs.
\end{abstract}



\section{Introduction} \label{sec:intro}

In the last few years, a large number of discs have revealed substructure in the form of rings and gaps when observed at millimeter wavelengths with ALMA \citep{2015ALMA,2016Andrews,2018Fedele,2018Andrews,2018Huang,2018Dipierro,2020Booth}. A few of these discs are thought to be very young ($<1$Myr) \citep{2015ALMA,2018Fedele,2018Dipierro}.  Ring and gap structures have also been observed in even younger ($\,\lesssim 0.5\,$Myrs) Class 1 discs \citep{2018Sheehan,Segura-Cox2020}.
Young discs are thought to be massive and could potentially be gravitationally unstable. Such discs are expected to harbour spiral arms.  There is evidence that discs with spiral arms in the midplane exist \citep{2016Perez, 2018Huang}, although they seem to be quite rare. Is the observed rarity due to young discs being less massive or can spiral structures in massive discs be hidden by other processes?

The goal of this Letter is to investigate the impact that a migrating giant planet, irrespective of how it formed, has on the structure of a gravitationally unstable disc and the resulting implications for observations of such discs. We find that a migrating giant planet is able to suppress spiral structures yielding discs that appear axisymmetric with rings and gaps. 

\section{Method} \label{sec:model}

\subsection{Hydrodynamical simulations \& initial conditions}

We perform scale free 3D gas hydrodynamic simulations using \textsc{phantom}, a smoothed particle hydrodynamics (SPH) code developed by \cite{2018Price}.

The disc setup is identical to that in \cite{2020Rowther}. We model a disc using 2 million particles between $R_{in} = 1$ and $R_{out} = 25$ in code units with a disc-to-star mass ratio of 0.1. The central star, and the planet are modelled as sink particles \citep{1995Bate}. The accretion radius of the central star is set to be equal to the disc inner boundary, $R_{in}$. To maintain a roughly constant planet mass throughout the simulation, the accretion radius of the planet is limited by setting it to 0.001 in code units. This is ${\sim} 10$ times smaller than the minimum Hill radius for a $3M_{Jup}$ planet at $R_{in}$. The initial surface mass density is set as a smoothed power law and is given by
\begin{equation}
\Sigma = \Sigma_{0}  \left ( \frac{R}{R_{0}} \right)^{-1} f_{s},
\end{equation}
where $\Sigma_{0}$ is the surface mass density at $R=R_{0}=1 $ and ${f_{s} = 1-\sqrt{R_{in}/R}}$ is {the factor used to smooth the surface density at the inner boundary of the disc}. The initial temperature profile is expressed as a power law
\begin{equation}
T = T_{0} \left ( \frac{R}{R_{0}} \right)^{-0.5},
\end{equation}
where $T_{0}$ is set such that the disc aspect ratio ${H/R=0.05}$ at $R=R_{0}$. The energy equation is 
\begin{equation}
\frac{\mathrm{d}u}{\mathrm{d}t} = -\frac{P}{\rho} \left ( \nabla \cdot \mathbf{v} \right) + \Lambda_{\mathrm{shock}} - \frac{\Lambda_{\mathrm{cool}}}{\rho}
\end{equation}
where we assume an adiabatic equation of state, $u$ is the specific internal energy, the first term on the RHS is the $P\mathrm{d}V$ work, $\Lambda_{\mathrm{shock}}$ is a heating term that is due to the artificial viscosity used to correctly deal with shock fronts, and 
\begin{equation}
\Lambda_{\mathrm{cool}} = \frac{\rho u}{t_{\mathrm{cool}}}
\end{equation}
controls the cooling in the disc. Here we use a simple implementation of the cooling time which is proportional to the dynamical time by a factor of $\beta$,
\begin{equation}
t_{\mathrm{cool}} = \beta(R)\Omega^{-1} = \beta_{0} \left( \frac{R}{R_{0}} \right)^{-2} \Omega^{-1},
\end{equation}
where $\Omega$ is the orbital frequency and we have varied $\beta$ with radius \citep{2020Rowther}. This allows us to mimic a realistic self-gravitating disc that is only gravitationally unstable in the outer regions \citep{2005Rafikov, 2009Stamatellos, 2009Rice, 2009Clarke}.

To model shocks, we use an artificial viscosity switch that utilises the time derivative of the velocity divergence introduced by \cite{2010Cullen}. The artificial viscosity parameter $\alpha_{v}$ has a maximum of $\alpha_{\text{max}} = 1$ near the shock and a minimum of $\alpha_{\text{min}} = 0$ far away from the shock. The artificial viscosity coefficient $\beta_{v}$ is set to 2 (see \citealt{2018Price}).

\subsubsection{Embedding the planet}

The simulation is first run for 10 orbits to allow spiral structure in the disc to develop. We then split it into two simulations, one with an embedded planet and one which continues to evolve without a planet. The latter is a control simulation to compare the impact a migrating giant planet has on the spiral structure in a gravitationally unstable disc.
A  planet with a planet-to-star mass ratio of ${q=2.9 \times 10^{-3}}$, equivalent to $3M_{\mathrm{Jup}}$ in a $0.1M_{\odot}$ disc around a $1M_{\odot}$ central star is added at $R_{p}=20$.

The simulations are run for another 8 orbits, with a total simulation time of 18 orbits.

\subsection{Post processing of simulations}

The raw synthetic continuum images at $1.3$mm are created using \textsc{mcfost} \citep{2006Pinte, 2009Pinte}. We first scale the simulations such that $R_{out} = 200$\textsc{au}, and initial ${R_{p}=160}$\textsc{au}. We use $10^{8}$ photon packets on a Voronoi tesselation where each \textsc{mcfost} cell corresponds to an SPH particle. The luminosity of the star is calculated using an assumed mass of $1M_{\odot}$ and a 1Myr isochrone from \cite{2000Siess} which corresponds to a temperature of $T_{\star} = 4286$K. Since the gas surface density of the disc is quite large, the Stokes numbers in the disc are small enough ($<0.1$ for mm sized grains) that we assume that the dust is well coupled to gas. Therefore we assume that the dust distribution is identical to the gas distribution and a constant dust-to-gas ratio of 0.01. The dust sizes vary between 0.3 and 1000 $\mu$m and are distributed across 100 different sizes with a power-law exponent of -3.5. We assume all dust grains are made of astronomical silicates, and are spherical and homogeneous. We compute the dust properties using Mie theory. The disc is assumed to be at a distance of 140pc.

To create mock millimeter continuum observations, we use the ALMA Observation Support Tool \citep{2011Heywood}. We use integration times of 12, 30, 60, and 120 minutes at 230GHz (1.3mm) in Band 6 with the ALMA Cycle 8 C43-7 configuration. We assume a bandwidth of 7.5GHz and a precipitable water vapour level of 0.913mm. \texttt{CLEAN} images are created using natural weights resulting in a beam size of $0.107{}'' \times 0.124{}''$, or equivalently $15.0$\textsc{au} $\times 17.3$\textsc{au} at 140pc. As shown by \cite{2016Mayer}, different ALMA configurations can alter the detectability of features. Hence a similar set mock observations are created with the C43-6 configuration ($0.156{}'' \times 0.196{}''$) and with a disc inclined at $40^{\circ}$ using the C43-7 configuration.

Assuming a disc inclination and position angle of $40^{\circ}$, optically thick $^{12}$CO, and optically thinner $^{13}$C$^{16}$O CO-isotopologue channel maps are generated for the ${J=3-2}$ transition using a velocity resolution of $0.1$ km/s. We choose this transition as it can be observed with a good compromise between observation time and signal-to-noise ratio. The abundances for $^{12}$CO and $^{13}$C$^{16}$O are assumed to be a fraction $1\times 10^{-4}$ and $2\times 10^{-7}$ of the total disc mass (i.e. relative to $H_{2}$) respectively. We account for CO freeze-out at $T < 20$K, and photo-dissociation and photo-desorption in regions of high UV radiation (see Appendix B of \citealt{2018Pinte}). We then convolve the resulting images with a beam size of $0.05{}'' \times 0.05{}''$; the maximum resolution expected for CO line observations.

\section{Results} \label{sec:results}

\subsection{Impact on spiral structure}

As shown in \cite{2020Rowther}, planets migrate inwards rapidly until they reach the gravitationally stable inner disc. For the simulations presented here, the gravitationally stable region is inside $R\approx120$\textsc{au}. The impact of the planet on the disc as it migrates is compared to the control simulation using the Toomre parameter \citep{1964Toomre},
\begin{equation}
Q = \frac{c_{s}\Omega}{\pi G \Sigma},
\end{equation}
which gives a measure of how gravitationally unstable the disc is, where $\Sigma$ and $c_{s}$ are the disc surface density and sound speed, respectively.

\begin{figure}
\begin{center}
\includegraphics[width=\linewidth]{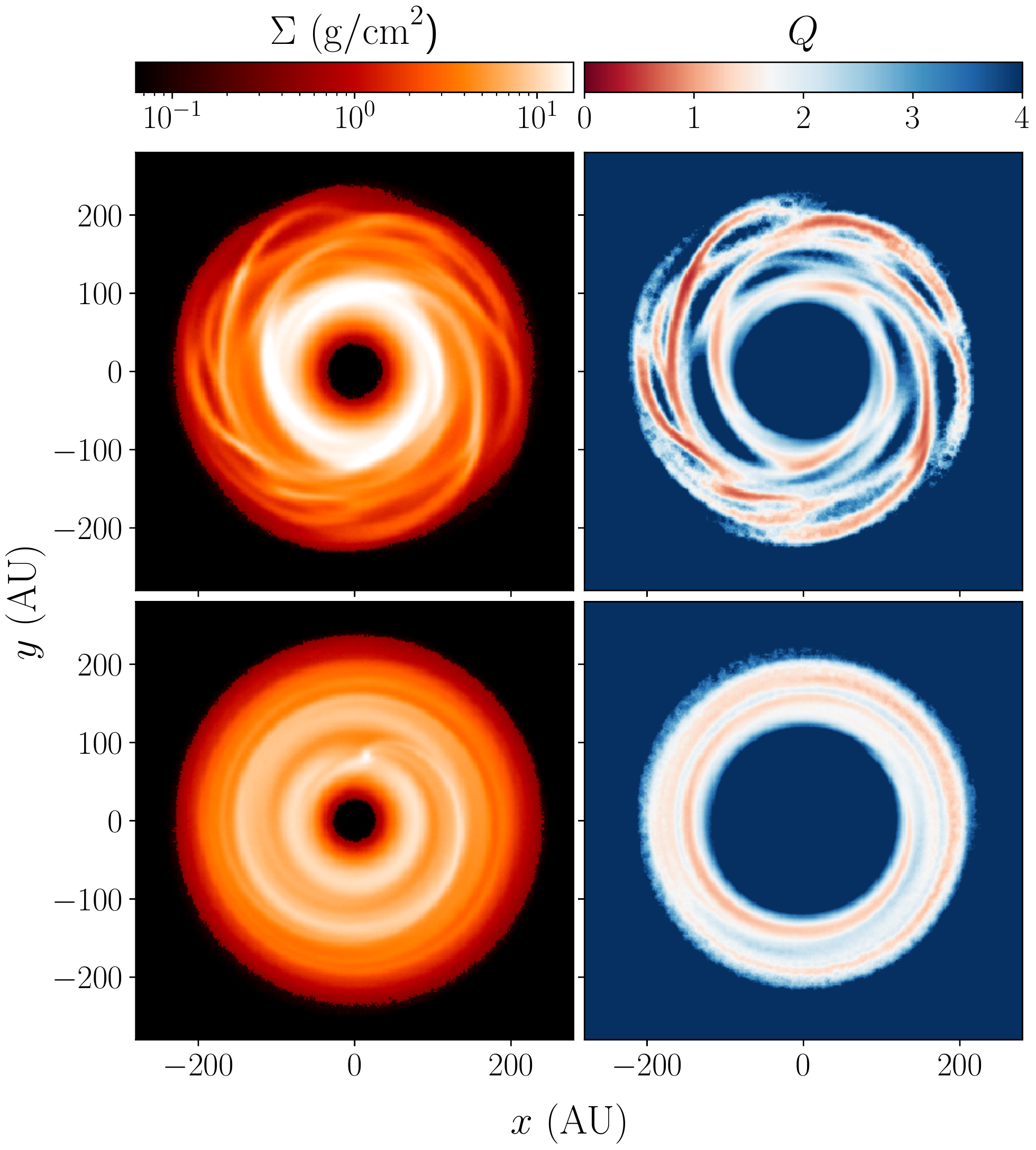}
\caption{Surface density (left) and 2D Toomre parameter (right) plots of a disc without a planet (top) and with a $3M_{\mathrm{Jup}}$ planet (bottom) at the end of the simulation.  The 2D Toomre plots show how gravitationally unstable the disc is. The critical value for non-axisymmetric instabilities is when $Q\lesssim1.7$ \citep{2007Durisen}, shown in red. The planet erases the spiral structure and is massive enough to carve out a gap. This results in a gravitationally stable axisymmetric disc.}
\label{Q2D_5J}
\end{center}
\end{figure}

Figure \ref{Q2D_5J} shows the surface density (left) and the 2D Toomre parameter $Q$ (right) of the simulations without a planet (top) and with a $3M_{\mathrm{Jup}}$ planet (bottom) at the end of the simulation. It is clear that the presence of the planet significantly impacts the structure. The planet suppresses the spiral structure and opens up a gap. 
Despite both discs retaining the same disc mass (${\sim}0.09M_{\odot}$) within 200\textsc{au}, the 2D $Q$ plots show that the presence of the planet results in the disc becoming mostly gravitationally stable with $Q>1.7$ throughout the disc. Without a planet, strong gravitationally unstable ($Q<1.7$) arms are present.

\begin{figure}
\begin{center}
\includegraphics[width=\linewidth]{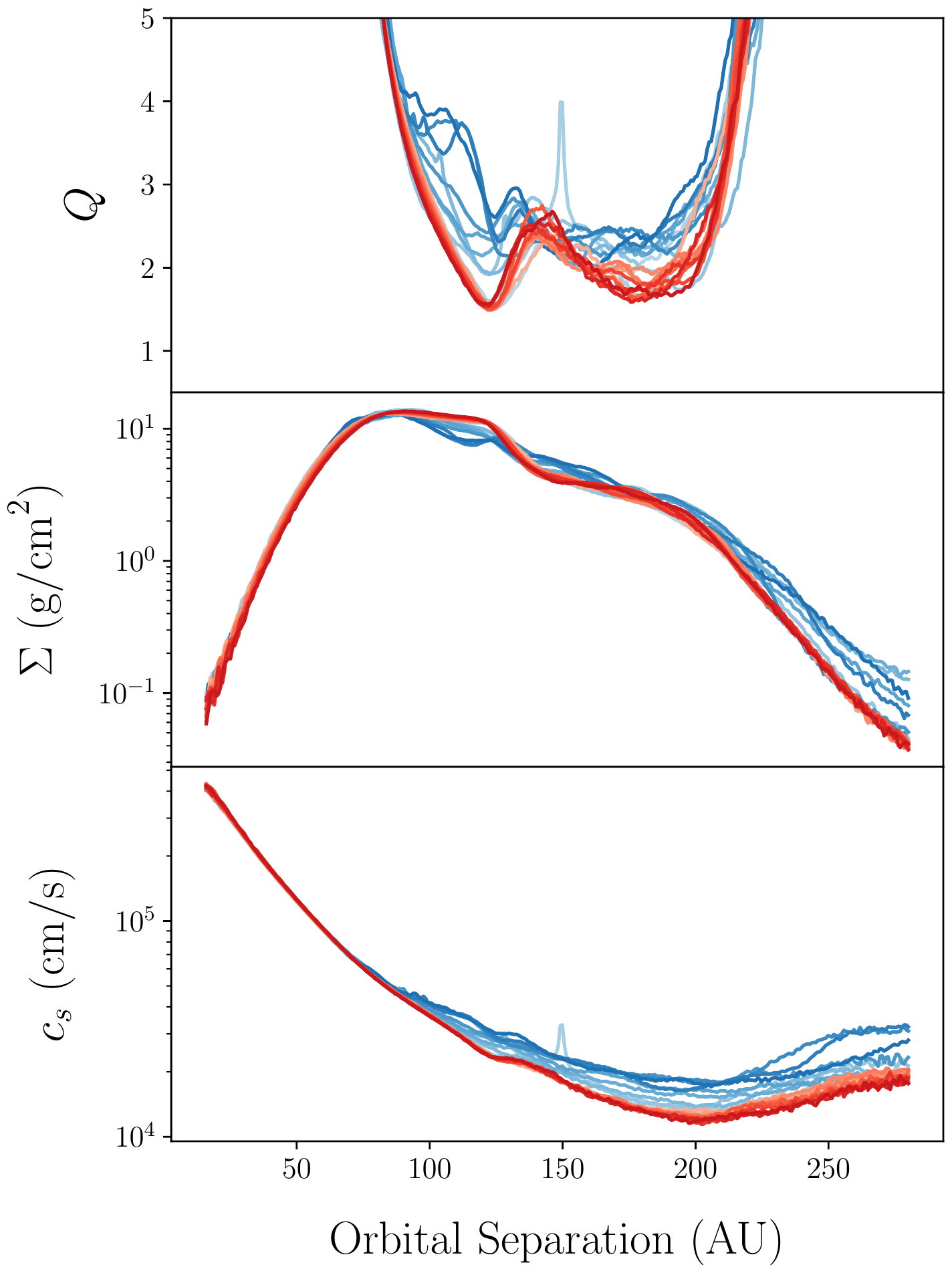}
\caption{Azimuthally averaged Toomre parameter, $Q$ (top), surface density, $\Sigma$ (middle), and the sound speed, $c_{s}$ (bottom) for both the disc without a planet (red lines) and with a $3M_{\mathrm{Jup}}$ planet (blue lines) for the first 2 orbits after the planet is added. The darker shades represent later times. The increase in $c_{s}$ due to the planet is the dominant reason for the increase in $Q$ in the outer regions of the disc (beyond $R\approx 120$\textsc{au}), causing the disc to become gravitationally stable.}
\label{csAzi}
\end{center}
\end{figure}

The dominant reason for the increase in $Q$ is due to the planet's influence on the disc temperature, or equivalently the sound speed $c_{s}$, in the disc. The spiral wakes generated by the planet as it migrates are regions of relative overdensities with respect to the disc background.  The radially propagating spiral wakes can evolve into shocks. The exchange of momentum and energy between the density wakes and the disc at the shocks influence the global properties of the disc, heating it up \citep{2001Goodman, Rafikov2016, 2020Ziampras}. 

The azimuthally averaged Toomre parameter (top), surface density (middle), and sound speed (bottom) is plotted in Figure \ref{csAzi} for the first 2 orbits after the $3M_{\mathrm{Jup}}$ planet (blue lines) is embedded. The red lines represent the disc without the planet at the same times. It can be seen, particularly in the regions of the disc beyond $R\approx 120$\textsc{au}, that there is a quick and significant increase in the sound speed after the planet has been added, whereas, the change in $\Sigma$ is relatively small. This increase in temperature causes the disc to become gravitationally stable, resulting in a higher value of the Toomre parameter.

To ensure the sudden inclusion of the $3M_{\mathrm{Jup}}$ planet did not trigger artificial excess heating, we compared the artificial viscosity to the gravitational stress parameter immediately after the planet was embedded. The increase in artificial viscosity due to the planet is negligible compared to the magnitude of the gravitational stress parameter.

\subsection{Continuum images}

\begin{figure*}
\begin{center}
\includegraphics[width= \textwidth]{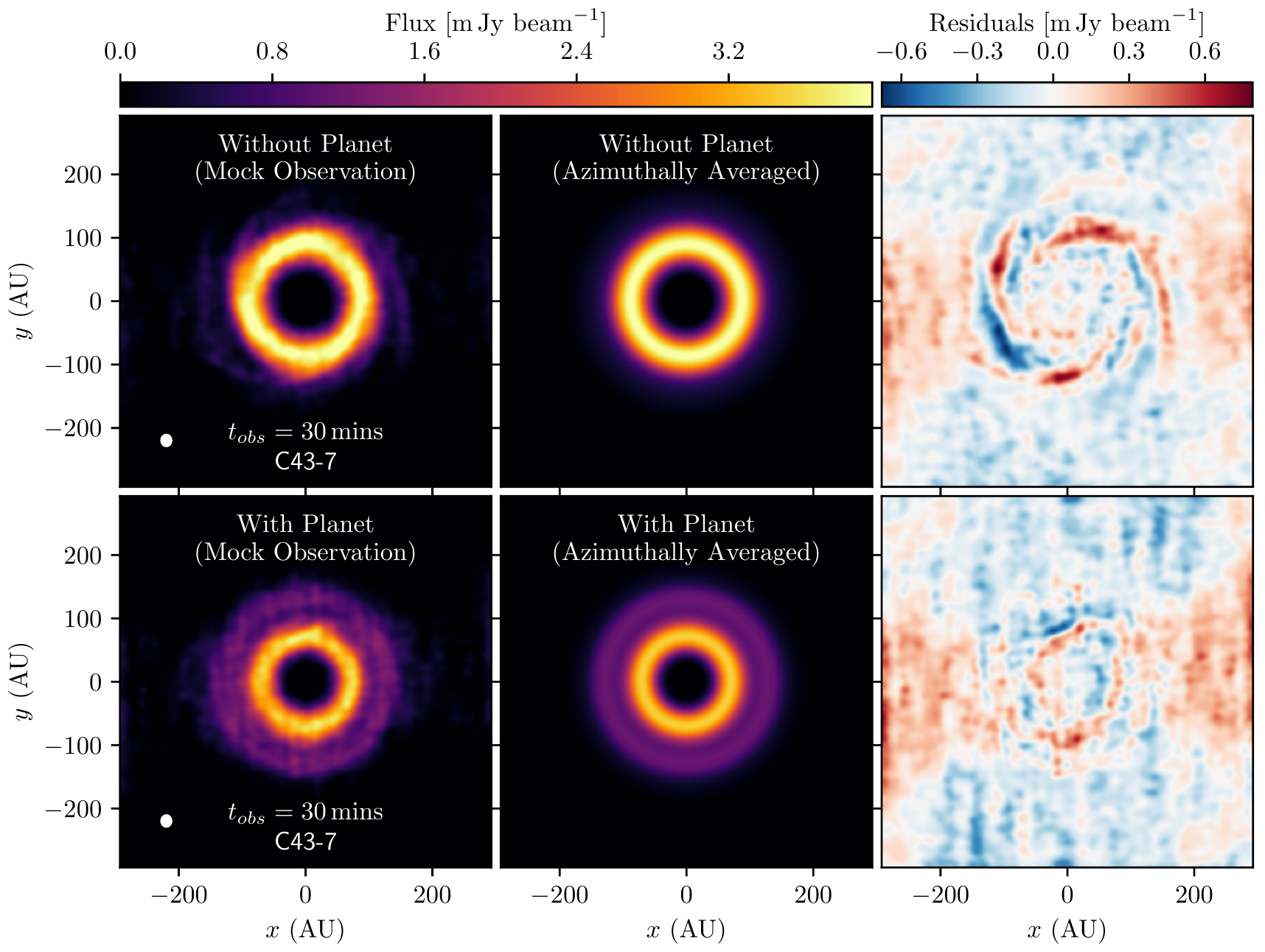}
\caption{Mock observation (left) using the C43-7 configuration  with an integration time of 30 minutes, axisymmetric flux map (middle), and residual flux (right) of a disc without a planet (top) and with a $3M_{\mathrm{Jup}}$ planet (bottom). The white ellipse in the bottom left of each mock observation represents the beam size. The presence of the planet results in a disc that no longer consists of spiral structure due to gravitational instability and appears to be more axisymmetric.}
\label{0pt5hr}
\end{center}
\end{figure*}

\begin{figure*}
\begin{center}
\includegraphics[width= 0.95\textwidth]{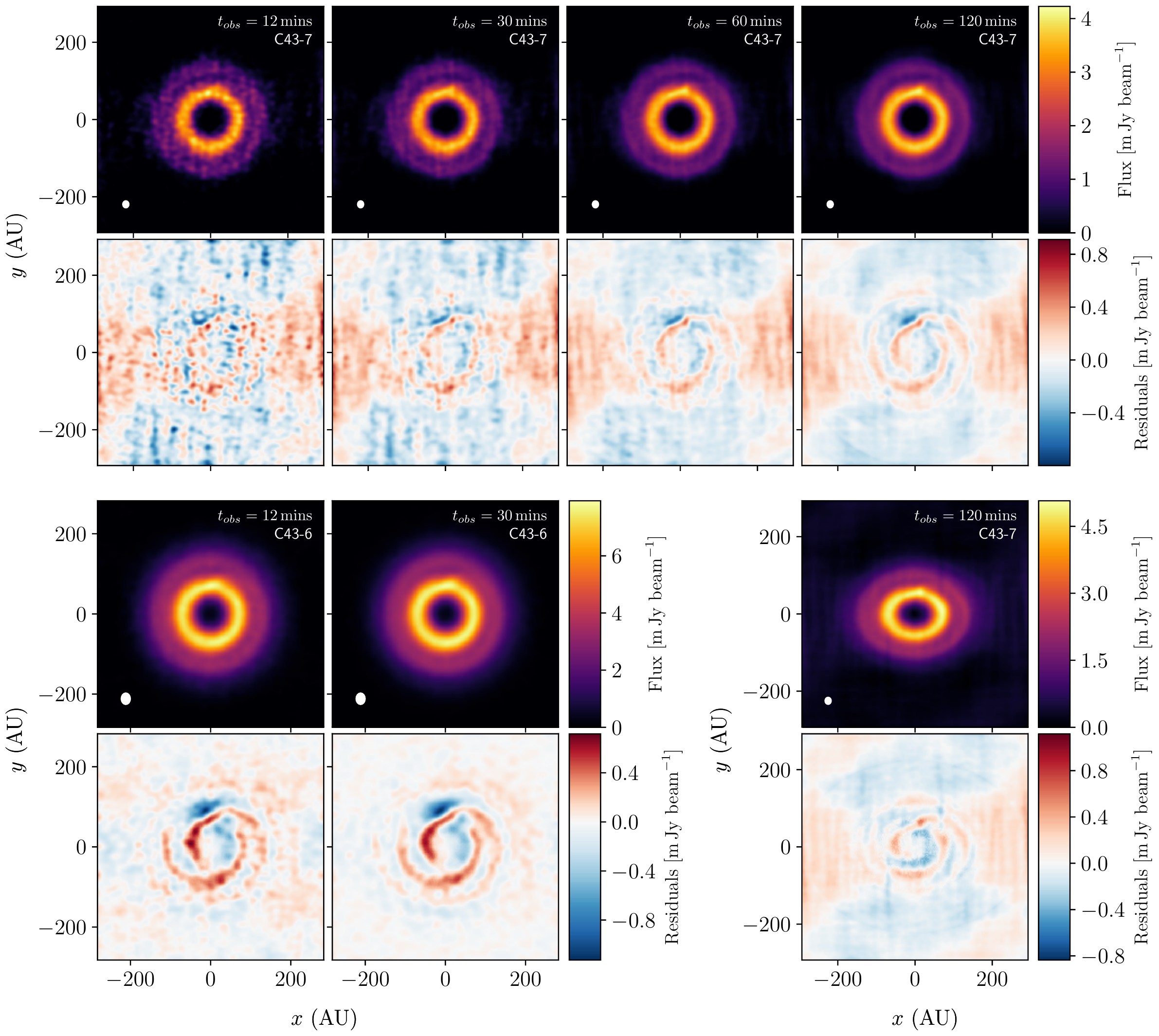}
\caption{Top subset of panels show the mock observations (top half) and residual plots (bottom half) of a disc with a $3M_{\mathrm{Jup}}$ planet with integration times of 12, 30, 60, and 120 minutes using the higher resolution C43-7 configuration. The bottom left subset of panels is similar but only for integrations times of 12 and 30 minutes using the C43-6 configuration. At high resolution the disc appears completely axisymmetric with short integration times. Whereas with longer integration times (over 60 minutes), spiral arms caused by the planet can be seen in the residual plots. At lower resolutions, the gap carved by the planet appears shallower, but the planet's spiral arms are easier to detect even with short integration times. The bottom right subset of panels show that the planet's spiral arms are also detectable for moderately inclined discs. This is shown for a disc inclined by $40^{\circ}$.}
\label{3JGallery}
\end{center}
\end{figure*}

The left column of Figure \ref{0pt5hr} compares the mock observations of a disc with a $3M_{\mathrm{Jup}}$ planet (bottom) and without a planet (top) with an integration time of 30 minutes using the C43-7 configuration. An axisymmetric flux map is produced as the azimuthally averaged flux of the mock observations and is shown in the middle column. Finally to highlight non-axisymmetric features, the residual flux is plotted in the right column by subtracting the axisymmetric map from the mock observation. 
The top subset of panels in Figure \ref{3JGallery} show the mock observations (top half) and residuals (bottom half) using the higher resolution C43-7 configuration. The bottom left subset of panels show the same with the lower resolution C43-6 configuration. The bottom right subset of panels are for a disc inclined by $40^{\circ}$ using the C43-7 configuration.

From Figure \ref{0pt5hr}, it can be seen that without a planet, spiral arms due to gravitational instability would be readily apparent. Whereas with the planet, the only major non-axisymmetric feature that remains are the spiral arms caused by the planet as shown in Figures \ref{0pt5hr} and \ref{3JGallery}. 
The higher resolution mock observations are able to better resolve the gap, however the planet induced spiral arms are only visible with integration times $>1$hr. This is true for both the non-inclined and moderately inclined disc. Using a lower resolution, which increases the signal-to-noise, allows the spiral arms to be easily seen in the residuals at lower integration times. The tradeoff is a gap that is less resolved and shallower in the continuum image. Despite the lower resolution images showing the spiral arms of the planet more easily, higher resolution images would still be favourable as a less resolved and shallower gap can lead to underestimates of the planet's mass. This could in turn lead to predictions with other methods of estimating the planet's mass, such as with the CO kinematics, that are inconsistent.

\subsection{CO kinematics}

\begin{figure*}
\centering
\gridline{\fig{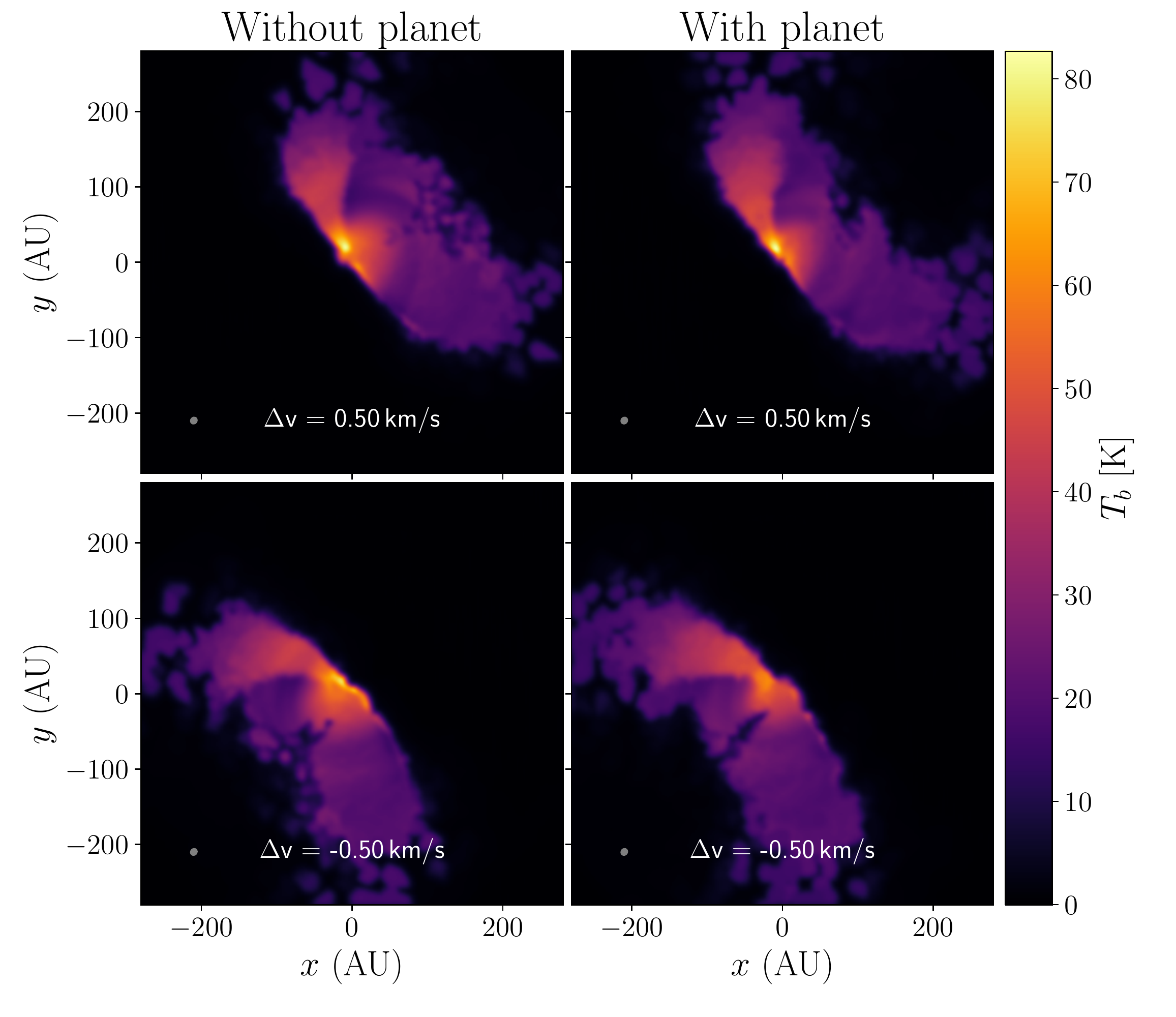}{0.49\textwidth}{(a) $^{12}$CO}
          \fig{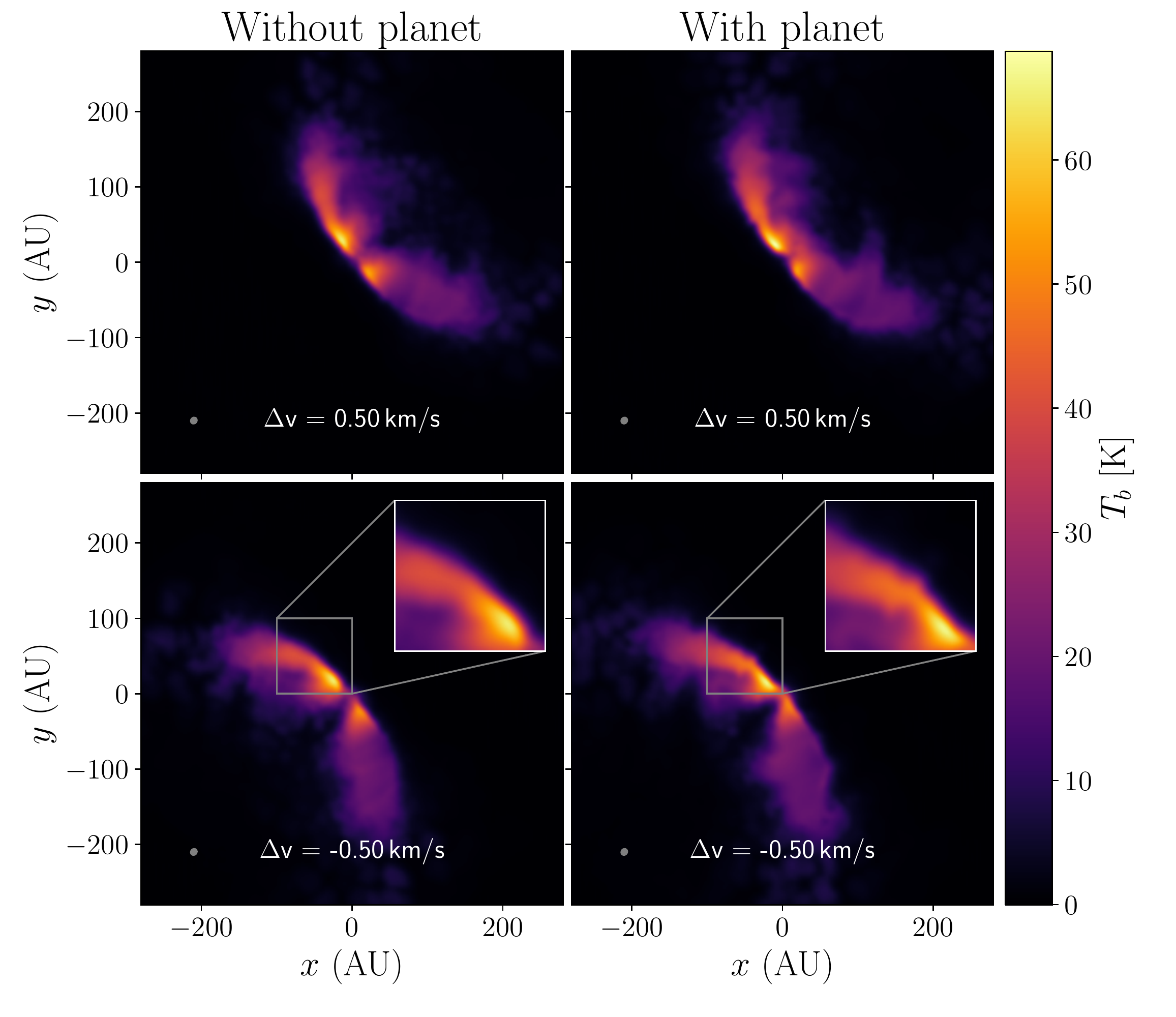}{0.49\textwidth}{(b) $^{13}$C$^{16}$O}}
\caption{Synthetic continuum subtracted channel maps ($\mathrm{J}=3-2$ transitions) for a) $^{12}$CO and b)  $^{13}$C$^{16}$O at $\Delta v = \pm 0.5 \, \mathrm{km \ s}^{-1}$ from the systemic velocity of the disc with and without the planet. A kink is not detected in the optically thick $^{12}$CO, but is visible in the optically thin $^{13}$C$^{16}$O in the negative velocity channel within the vicinity of the planet, see bottom right inset in Fig \ref{CO}b. The bend in the velocity profile in the bottom right inset is referred to as the kink. This is contrasted with the smooth velocity profile in the bottom left inset. The disc inclination and position angle are both $40^{\circ}$. }
\label{CO}
\end{figure*}

The spiral waves generated by the planet can cause localised deviations in the Keplerian flow of the disc as shown recently by \cite{2019Pinte, 2020Pinte}. These deviations can be detected as \textit{kinks} in the gas channel maps.

Figure \ref{CO} shows the channel maps for  $^{12}$CO and $^{13}$C$^{16}$O at $\Delta v = \pm 0.5 \, \mathrm{km \ s}^{-1}$ from the systemic velocity with the continuum subtracted. The optically thick $^{12}$CO does not reveal any signs of a planet, but a kink is noticeable in the optically thinner $^{13}$C$^{16}$O in the $ -0.5 \, \mathrm{km \ s}^{-1}$ channel within the vicinity of the planet. We can exclude large scale perturbations or any azimuthally symmetric mechanisms since the kink is not seen in the opposite velocity channel where the profile remains smooth.  The kink is shown in the bottom right panel of Fig \ref{CO}b in the inset. The kink is also seen in the channels from $v = -0.3  \mathrm{ \ to \ } {-1.2}\, \mathrm{km \ s}^{-1}$. 

Although the channel maps without the planet are not perfectly smooth as would be expected from a Keplerian disc, this is likely to be due to the gravitational instabilities. \cite{2020Hall} show that for a gravitationally unstable disc the perturbations due to gravitational instabilities are seen in all azimuths and in both the positive and negative velocity channels. However in our simulations with a planet, the kink is more localised appearing in only the negative velocity channel.
To ensure that the observed kink is due to the planet, we re-calculated the channel maps with the planet at slightly different times with the planet being in different azimuthal locations. Although the kink is not detectable at all times, whenever the kink is detected it is in a different location determined by the position of the planet. Additionally, in some cases the kink was only detected for a narrower range of $\Delta v$.

\section{Discussion} \label{sec:discussion}

\subsection{Implications for dust-to-gas mass ratios}

In the early stages of evolution, discs are expected to be massive and gravitationally unstable \citep[e.g.][]{2018Bate}. A characteristic feature of such discs are their spiral arms. However, from observations so far \citep{2016Perez, 2018Huang}, these discs appear to be quite rare. Due to the difficulty in directly measuring the gas mass of the disc, the disc mass is often inferred via the dust mass using a fixed dust-to-gas mass ratio. In some scenarios, such as with MWC 480 \citep{2018Liu}, the observed dust mass can be high enough such that inferring the gas mass via the canonical dust-to-gas mass ratio of 0.01 results in a disc that is massive enough to be gravitationally unstable. Thus in their models, they assume a higher dust-to-gas mass ratio to ensure that the disc is gravitationally stable.
\cite{2020Facchini}  show that LkCa-15 retains a significant amount of dust mass in its rings. Consequently, when modelling \text{LkCa-15}, they assume an upper limit on the disc gas mass (and hence a higher dust-to-gas mass ratio) to ensure the disc is not locally gravitationally unstable.

In both cases this assumption was deemed necessary to explain the lack of spiral features in the observations, which would be expected from more massive discs.
However, we show that a migrating $3M_{\mathrm{Jup}}$ planet can cause a massive disc to become gravitationally stable and suppress any spiral structure that would otherwise be present. Hence, large dust-to-gas mass ratios are not necessarily required to explain a lack of spiral features.

\subsection{Caveats}

A caveat to the results presented here is the long term gravitational stability of the discs due to the cooling mechanism used. Although it mimics the characteristics of a realistic self-gravitating disc, i.e. a disc which is only gravitationally unstable in the outer regions, it is still a straight-forward implementation where the cooling time is a simple expression determined by the location in the disc. 

The actual cooling in the disc is likely to be more complex and evolve over time. Recently \cite{2018Mercer} calculated an effective $\beta$ from their radiative transfer calculations, which was found to vary both spatially and with time. Hence, it is unknown what impact using more complex cooling methods such as radiative transfer to cool the disc will have on the long term gravitational stability of the disc with a migrating planet. Whilst we show that a migrating giant planet can erase the spiral structure in a self-gravitating disc, it remains to be investigated whether the loss of spiral structure remains for a significant amount of time.

The results presented here initially appear contrary to \cite{2015Meru} where it was shown that a fragment formed by gravitational instability in the outer disc could trigger subsequent fragmentation in the inner disc. The main difference is that the Toomre profiles are quite different. In \cite{2015Meru}, the inner disc was on the edge of fragmentation ($Q \sim 1$). Hence the increase in the surface density as a result of the fragment driving material inwards towards the disc caused the disc to fragment. This behaviour is not seen or expected in this work as the inner disc remains well above the gravitationally stable regime ($Q \gtrsim 1.7)$.

In this work, we have only presented results for one planet and disc mass. 
Lower mass planets would be less likely to impact self-gravitating spiral structures. On the other hand, lower mass discs, which have weaker gravitational instabilities, would be more susceptible to having their spiral structures suppressed by planets. Although we find that a planet that is large enough to open up a gap also suppresses the spiral structure, we cannot conclude that non-gap-opening planets are unable to affect the disc structure.
We will present a follow up study that considers a variety of planet and disc properties, as well as ALMA configurations, to investigate the conditions at which spiral suppression is likely to be observed.

\section{Conclusion}
\label{sec:conc}

We perform 3D SPH simulations to investigate the impact a migrating giant planet has on the structure of a gravitationally unstable disc. Our work shows that the presence of the planet suppresses the spiral structure in the disc and causes the disc to become gravitationally stable because it alters the temperature structure.  This interaction between the planet and the disc causes the self-gravitating phase of the disc to be shortened, while retaining the same disc mass. The planet is able to carve open a gap resulting in an axisymmetric disc. 

The mock ALMA observations of the continuum presented here show that  the disc can appear completely axisymmetric for higher resolution ALMA configurations. However, with longer integration times or by sacrificing resolution, spiral arms from the planet become observable. In the latter case, the gap carved by the planet will be less resolved and appear shallower.  We also show that the planet can be detected with high resolution kinematics using optically thin CO-isotopologues like $^{13}$C$^{16}$O. Our results show it is possible to explain a lack of spiral structure in high mass discs without requiring high dust-to-gas mass ratios to limit the gas mass.

\section*{acknowledgments}

We thank the anonymous referee for their useful comments which benefited this work. SR acknowledges support from the Royal Society Enhancement Award. FM acknowledges support from the Royal Society Dorothy Hodgkin Fellowship. GMK is supported by the Royal Society as a Royal Society University Research Fellow. RN acknowledges funding from the European Research Council (ERC) under the European Union's Horizon 2020 research and innovation programme (grant agreement No 681601). This work was performed using Orac and Tinis HPC clusters at the University of Warwick.

\software{Matplotlib \citep{Hunter:2007},
numpy \citep{5725236},
astropy \citep{2018:Astropy},
\textsc{phantom} \citep{2018Price},
 \textsc{splash} \citep{2007Price},
\textsc{mcfost} \citep{2006Pinte, 2009Pinte}
}

\bibliography{HidingGravitationalInstability.bib}{}
\bibliographystyle{aasjournal}

\end{document}